\documentclass[12pt]{article}
\usepackage{amssymb}

\renewcommand{\theequation}{\arabic{section}.\arabic{equation}}

\newcommand{\xb}{\mbox{\boldmath $x$}}
\newcommand{\etap}{\eta^{\dagger}}
\newcommand{\half}{\frac{1}{2}}
\newcommand{\ca}{{\cal A}}
\newcommand{\tca}{\tilde{\cal A}}
\def\sech{\mathop{\rm sech}\nolimits}
\def\csch{\mathop{\rm csch}\nolimits}
\def\deg{\mathop{\rm deg}\nolimits}

\title{
First-order Intertwining Operators and Position-dependent Mass Schr\"odinger Equations
in d Dimensions}
\author{C. Quesne\\ 
{\sl \small Physique Nucl\'eaire Th\'eorique et Physique
Math\'ematique,  Universit\'e Libre de Bruxelles,} \\ 
{\sl \small Campus de la Plaine CP229, Boulevard~du Triomphe, B-1050 Brussels,
Belgium}\\
{\small E-mail: cquesne@ulb.ac.be}}
\date{ }
\begin{document}
\baselineskip=20pt plus 1pt minus 1pt
\maketitle

\begin{abstract}
The problem of $d$-dimensional Schr\"odinger equations with a position-dependent mass
is analyzed in the framework of first-order intertwining operators. With the pair $(H,
H_1)$ of intertwined Hamiltonians one can associate another pair of second-order partial
differential operators $(R, R_1)$, related to the same intertwining operator and such
that $H$ (resp.\ $H_1$) commutes with $R$ (resp.\ $R_1$). This property is
interpreted in superalgebraic terms in the context of supersymmetric quantum
mechanics (SUSYQM). In the two-dimensional case, a solution to the resulting system of
partial differential equations is obtained and used to build a physically-relevant model
depicting a particle moving in a semi-infinite layer. Such a model is solved by employing
either the commutativity of $H$ with some second-order partial differential operator $L$
and the resulting separability of the Schr\"odinger equation or that of $H$ and $R$
together with SUSYQM and shape-invariance techniques. The relation between both
approaches is also studied. 
\end{abstract}

\vspace{0.5cm}



\newpage
%
%
\section{INTRODUCTION}

The concept of position-dependent mass (PDM) is known to play an important role in the
energy-density functional approach to the quantum many-body problem in the context of
nonlocal terms of the accompanying potential. This formalism has been extensively used
in nuclei~\cite{ring}, quantum liquids~\cite{arias}, $^3$He clusters~\cite{barranco}, and
metal clusters~\cite{puente}.\par
%
%
Another area wherein the PDM approximation provides a very useful tool is the study of
electronic properties of many condensed-matter systems, such as
semiconductors~\cite{bastard} and quantum dots~\cite{serra}. In particular, recent
progress in crystal-growth techniques (molecular-beam-epitaxy technique, for instance)
for producing nonuniform semiconductor specimens, wherein the carrier effective mass
depends on position, has considerably enhanced the interest in the theoretical description
of semiconductor heterostructures.\par
%
%
{}Furthermore, PDM presence in quantum mechanical problems may reflect some other
unconventional effects, such as a deformation of the canonical commutation relations or
a curvature of the underlying space~\cite{cq}. It has also recently been signalled in the
rapidly-growing field of PT-symmetric~\cite{bender} (or, more generally,
pseudo-Hermitian~\cite{mosta}) quantum mechanics as occurring in the Hermitian
Hamiltonian equivalent to the PT-symmetric cubic anharmonic oscillator at lowest order of
perturbation theory~\cite{jones}.\par
%
%
All these developments have stimulated the search for exact solutions of PDM quantum
mechanical problems both in the nonrelativistic~\cite{cq, dekar, milanovic, plastino,
dutra, roy, koc, alhaidari02, gonul, bagchi04, bagchi05a, bagchi05b, yu, chen, dong} and
relativistic~\cite{alhaidari04, vakarchuk} contexts since they may provide a conceptual
understanding of some physical phenomena, as well as a testing ground for some
approximation schemes. In the nonrelativistic case, which we are going to consider here,
the generation of PDM and potential pairs leading to exactly solvable, quasi-exactly
solvable or conditionally-exactly solvable Schr\"odinger equations has been achieved by
extending some methods known in the constant-mass case, such as point canonical
transformations~\cite{bhatta}, Lie algebraic methods~\cite{alhassid}, as well as
supersymmetric quantum mechanical (SUSYQM) and shape-invariance 
techniques~\cite{cooper}.\par
%
%
Most of these works have been devoted to one-dimensional systems. The aim of the
present paper is to tackle the more difficult problem of generating exact solutions for
$d$-dimensional PDM Schr\"odinger equations. For such a purpose, we will extend an
approach that has proved very useful in one dimension, namely that of intertwining
operators~\cite{bagchi04}, related to SUSYQM methods (see also \cite{cq, milanovic,
plastino, dutra, roy, koc, gonul, bagchi05a, bagchi05b}). As a first step, we plan to
consider here the case of first-order intertwining operators and their SUSYQM
interpretation.\par
%
%
In Section~2, the general theory of $d$-dimensional PDM Hamiltonians admitting a
first-order intertwining operator is developed and a solution to the resulting system of
partial differential equations is obtained in the two-dimensional case. From this solution, a
new exactly-solvable PDM model in a semi-infinite layer is built in Section~3. In
Subsections~3.1 and 3.2, the separability of the corresponding Schr\"odinger equation
and the existence of an intertwining operator are shown to provide two alternative
methods for getting the bound-state spectrum and wavefunctions. The relation between
both approaches is then reviewed in Subsection~3.3. Section~4 contains some final
remarks.\par
%
%
\section{GENERAL THEORY AND SUPERSYMMETRIC INTERPRETATION}

On using the von Roos general two-parameter form of the PDM kinetic energy
operator~\cite{vonroos}, choosing units wherein $\hbar = 2m_0 = 1$ and summing over
dummy indices, the $d$-dimensional PDM Schr\"odinger equation may be written as
\begin{equation}
  \left\{- \half \left[M^{\alpha}(\xb) \partial_i M^{\beta}(\xb) \partial_i
  M^{\gamma}(\xb) + M^{\gamma}(\xb) \partial_i M^{\beta}(\xb) \partial_i
  M^{\alpha}(\xb)\right] + V(\xb)\right\} \psi(\xb) = E \psi(\xb).  \label{eq:schrodinger} 
\end{equation}
Here $\xb \equiv (x_1, x_2, \ldots, x_d)$, $\partial_i \equiv \partial/\partial x_i$,
$i=1$, 2,~\ldots, $d$, $M(\xb)$ is the dimensionless form of the mass function
$m(\xb) = m_0 M(\xb)$, $V(\xb)$ is the potential and $\alpha$, $\beta$, $\gamma$ are
the von Roos ambiguity parameters, constrained by the condition $\alpha + \beta +
\gamma = -1$. The expression of the kinetic energy operator in (\ref{eq:schrodinger})
has the advantage of having an inbuilt Hermiticity and of containing  as special cases all
the proposals made in the literature to cope with noncommutativity of the momentum and
PDM operators.\par
%
%
As is well known (see, e.g., \cite{cq, bagchi04}), we can get rid of the ambiguity
parameters in the kinetic energy operator by transferring them to the effective potential
energy of the variable mass system. Equation (\ref{eq:schrodinger}) therefore acquires
the form
\begin{equation}
  H \psi(\xb) \equiv \left(- \partial_i \frac{1}{M(\xb)} \partial_i + V_{\rm eff}(\xb)\right)
  \psi(\xb) = E \psi(\xb),  \label{eq:schrodinger-bis}
\end{equation}
where
\begin{equation}
  V_{\rm eff}(\xb) = V(\xb) + \half (\beta+1) \frac{\Delta M}{M^2} - [\alpha
  (\alpha+\beta+1) + \beta + 1] \frac{(\partial_i M)(\partial_i M)}{M^3}
\end{equation}
and $\Delta \equiv \partial_i \partial_i$ denotes the $d$-dimensional Laplacian
operator.\par
%
%
In the following, we shall be interested in the bound-state energy spectrum $E_n$,
$n=0$, 1, 2,~\ldots, of $H$ and in the corresponding wavefunctions $\psi_n(\xb)$,
$n=0$, 1, 2,~\ldots. To be physically acceptable, the latter should satisfy two
conditions~\cite{bagchi05a}:\par
\noindent (i) As in conventional quantum mechanics with a constant mass, they should be
square integrable over the domain $D$ on which $M(\xb)$ and $V(\xb)$ are defined, i.e.,
\begin{equation}
  \int_D d\xb\,  |\psi_n(\xb)|^2 < \infty.  \label{eq:wf-C1}
\end{equation}
\noindent (ii) To ensure that $H$ be Hermitian in the Hilbert space to which
the $\psi_n(\xb)$'s belong, they should be such that
\begin{equation}
  \frac{|\psi_n(\xb)|^2}{\sqrt{M(\xb)}} \to 0  \qquad \mbox{\rm on the boundary of
  $D$}. \label{eq:wf-C2}
\end{equation}
This extra condition may have relevant effects whenever $M(\xb)$ vanishes at some
boundary point.\par
%
%
Let us consider the intertwining relationship
\begin{equation}
  \eta H = H_1 \eta,  \label{eq:inter}
\end{equation}
where $H_1$ has the same kinetic energy term as $H$ and an associated effective
potential $V_{\rm 1, eff}(\xb)$. Choosing a first-order intertwining operator
\begin{equation}
  \eta = A^{(i)}(\xb) \partial_i + B(\xb),  \label{eq:eta}
\end{equation}
we are led to the restrictions
\begin{equation}
  \partial_i A^{(j)} + \partial_j A^{(i)} = - \delta_{i,j} A^{(k)} \frac{\partial_k M}{M},
  \label{eq:system-1}
\end{equation}
\begin{eqnarray}
  A^{(i)} (V_{\rm eff} - V_{\rm 1, eff}) & = & - A^{(j)} \left(\frac{\partial_i\partial_j M}
       {M^2} - 2 \frac{(\partial_i M)(\partial_j M)}{M^3}\right) - \frac{1}{M} \Delta A^{(i)}
        \nonumber \\
  && \mbox{} + \left(\partial_j A^{(i)}\right) \frac{\partial_j M}{M^2} - \frac{2}{M}
        \partial_i B,  \label{eq:system-2}
\end{eqnarray}
\begin{equation}
  A^{(i)} \partial_i V_{\rm eff} = - B (V_{\rm eff} - V_{\rm 1, eff}) - \frac{1}{M} \Delta
  B + (\partial_i B) \frac{\partial_i M}{M^2}.  \label{eq:system-3}
\end{equation}
\par
%
%
{}From the intertwining relation (\ref{eq:inter}) and the Hermiticity of $H$ and $H_1$, it
follows that
\begin{equation}
  H \etap = \etap H_1,
\end{equation}
showing that
\begin{equation}
  R \equiv \etap \eta, \qquad R_1 \equiv \eta \etap  \label{eq:R}
\end{equation}
are such that
\begin{equation}
  [H, R] = 0, \qquad [H_1, R_1] = 0.
\end{equation}
Hence $R$ is an integral of motion for the quantum mechanical problem described by
$H$ (provided it is independent of the latter). A similar situation occurs for the pair
$(H_1, R_1)$. Inserting $\eta$ from (\ref{eq:eta}) and $\etap = - A^{(i)} \partial_i -
(\partial_i A^{(i)}) + B$ in (\ref{eq:R}), we get
\begin{eqnarray}
  R & = & - A^{(i)} A^{(j)} \partial^2_{ij} - [A^{(i)} (\partial_j A^{(j)}) + A^{(j)}
       (\partial_j A^{(i)})] \partial_i - A^{(i)} (\partial_i B) - B (\partial_i A^{(i)}) + B^2,
       \nonumber \\
  R_1 & = & R - A^{(i)} [(\partial^2_{ij} A^{(j)}) - 2 (\partial_i B)].  \label{eq:R-bis} 
\end{eqnarray}
\par
%
%
The results obtained so far are amenable to a superalgebraic interpretation. Since
\begin{equation}
  \eta R = R_1 \eta, \qquad R \etap = \etap R_1,  \label{eq:inter-bis}
\end{equation}
we may indeed introduce some matrix operators
\begin{equation}
  {\cal H} = \left( \begin{array}{cc}
      H & 0 \\ 0 & H_1
      \end{array}\right), \qquad
  {\cal R} = \left( \begin{array}{cc}
      R & 0 \\ 0 & R_1
      \end{array}\right), \qquad
  {\cal Q}^+ = \left( \begin{array}{cc}
      0 & 0 \\ \eta & 0
      \end{array}\right), \qquad
  {\cal Q}^- = \left( \begin{array}{cc}
      0 & \etap \\ 0 & 0
      \end{array}\right),
\end{equation}
satisfying the defining relations of the gl(1/1) superalgebra~\cite{scheunert}, i.e.,
\begin{equation}
  \{{\cal Q}^+, {\cal Q}^-\} = 0, \qquad \{{\cal Q}^{\pm}, {\cal Q}^{\pm}\} = 0, \qquad
  [{\cal R}, {\cal Q}^{\pm}] = [{\cal H}, {\cal Q}^{\pm}] = [{\cal H}, {\cal R}] = 0. 
  \label{eq:gl(1/1)}
\end{equation}
In such a context, ${\cal R}$, ${\cal Q}^+$ and ${\cal Q}^-$ realize the usual SUSYQM
sl(1/1) superalgebra~\cite{cooper}, while $\cal H$ turns out to be the Casimir operator
of gl(1/1).\par
%
%
In the one-dimensional case, it has been shown elsewhere~\cite{bagchi04} that the
solution of Eqs.~(\ref{eq:system-1}) -- (\ref{eq:system-3}) can be written as
\begin{equation}
  A(x) = M^{-1/2}, \qquad V_{\rm eff}(x) = - (AB)' + B^2 + \lambda, \qquad 
  V_{\rm 1,eff}(x) = V_{\rm eff} - A (A'' - 2B'),  \label{eq:sol-1dim}
\end{equation}
where $\lambda$ is some integration constant. Note that from (\ref{eq:sol-1dim})
onwards, we denote by a prime the derivative of any function depending on a single
variable. On combining now (\ref{eq:sol-1dim}) with (\ref{eq:R-bis}), we arrive at the
trivial outcome
\begin{equation}
  R = H - \lambda, \qquad R_1 = H_1 - \lambda.
\end{equation}
\par
%
%
To get significant results in (\ref{eq:R}) -- (\ref{eq:gl(1/1)}), we must therefore be at
least in two dimensions. In such a case, the existence of the integral of motion $R$
(resp.\ $R_1$) means that $H$ (resp.\ $H_1$) is integrable. The first constraint
(\ref{eq:system-1}) can now be written as
\begin{equation}
  \partial_x A^{(1)} = - \frac{1}{2M} (A^{(1)} \partial_x M + A^{(2)} \partial_y M),
  \qquad  \partial_y A^{(1)} = - \partial_x A^{(2)}, \qquad \partial_x A^{(1)} =
  \partial_y A^{(2)},  \label{eq:system-1-bis} 
\end{equation}
where we henceforth denote $x_1$, $x_2$, $\partial_1$, $\partial_2$ by $x$, $y$,
$\partial_x$, $\partial_y$, respectively. Conditions (\ref{eq:system-1-bis}) imply that
\begin{equation}
  \Delta A^{(1)} = \Delta A^{(2)} = 0,
\end{equation}
thus leading to a small simplification in the second constraint (\ref{eq:system-2}).\par
%
%
Let us find the general solution to Eq.~(\ref{eq:system-1-bis}) in the special case where
the PDM depends on a single variable, e.g.,
\begin{equation}
  M = M(x).  \label{eq:M(x)}
\end{equation}
The first constraint in (\ref{eq:system-1-bis}) then amounts to
\begin{equation}
  \frac{\partial_x A^{(1)}}{A^{(1)}} = - \frac{M'}{2M},
\end{equation}
so that
\begin{equation}
  A^{(1)} = \frac{f(y)}{\sqrt{M}},  \label{eq:A1}
\end{equation}
where $f(y)$ is a so far undetermined function of $y$. Plugging (\ref{eq:A1}) in the
second constraint of (\ref{eq:system-1-bis}), we find
\begin{equation}
  A^{(2)} = - f'(y) \int^x \frac{dx'}{\sqrt{M(x')}} + g(y)
\end{equation}
in terms of another undetermined function $g(y)$. It remains to solve the third
constraint in (\ref{eq:system-1-bis}), which can be rewritten as
\begin{equation}
  f(y) \left(\frac{1}{\sqrt{M}}\right)' = - f''(y) \int^x \frac{dx'}{\sqrt{M(x')}} + g'(y).
  \label{eq:constraint3}
\end{equation}
On deriving both sides of this relation with respect to $x$, we arrive at the conclusion
that
\begin{equation}
  \frac{f''(y)}{f(y)} = - \frac{\left(1/\sqrt{M(x)}\right)''}{1/\sqrt{M(x)}} = C,  \label{eq:C}
\end{equation}
where $C$ is some constant. Let us now distinguish between the three subcases $C<0$,
$C=0$ and $C>0$.\par
%
%
Whenever $C = - q^2 < 0$ (where we may assume $q>0$), we find from (\ref{eq:C})
\begin{equation}
  f(y) = a \sin qy + b \cos qy, \qquad \frac{1}{\sqrt{M}} = c \sinh qx + d \cosh qx,
\end{equation}
where $a$, $b$, $c$, $d$ are four integration constants. Inserting these relations in
(\ref{eq:constraint3}) shows that $g(y)$ must reduce to some constant $g$. We
conclude that in the first subcase, the general solution of (\ref{eq:system-1-bis}),
corresponding to the choice (\ref{eq:M(x)}), is given by
\begin{eqnarray}
  M(x) & = & \frac{1}{(c \sinh qx + d \cosh qx)^2}, \nonumber \\
  A^{(1)}(x,y) & = & (a \sin qy + b \cos qy) (c \sinh qx + d \cosh qx), \nonumber \\
  A^{(2)}(x,y) & = & (- a \cos qy + b \sin qy) (c \cosh qx + d \sinh qx) + g.
      \label{eq:C-negative}
\end{eqnarray}
\par
%
%
By proceeding in a similar way for $C = 0$ and $C = q^2$ (with $q>0$), we respectively
obtain
\begin{eqnarray}
  M(x) & = & \frac{1}{(cx+d)^2},  \nonumber \\
  A^{(1)}(x,y) & = & (ay+b) (cx+d),  \nonumber \\
  A^{(2)}(x,y) & = & - \half ac (x^2-y^2) - adx +bcy + g,  \label{eq:C-null}
\end{eqnarray}
and
\begin{eqnarray}
  M(x) & = & \frac{1}{(c \sin qx + d \cos qx)^2}, \nonumber \\
  A^{(1)}(x,y) & = & (a \sinh qy + b \cosh qy) (c \sin qx + d \cos qx), \nonumber \\
  A^{(2)}(x,y) & = & (a \cosh qy + b \sinh qy) (c \cos qx - d \sin qx) + g,  \label{eq:C-positive}
\end{eqnarray}
in terms of five arbitrary constants $a$, $b$, $c$, $d$, $g$.\par
%
%
Considering next Eq.~(\ref{eq:system-2}) in the case where $M$ is given by
(\ref{eq:M(x)}), we get
\begin{eqnarray}
  A^{(1)}(V_{\rm eff} - V_{\rm 1, eff}) & = & - \frac{2C}{M} A^{(1)} - \frac{2}{M}
        \partial_x B,  \nonumber \\
  A^{(2)}(V_{\rm eff} - V_{\rm 1, eff}) & = & - \left(\frac{1}{M}\right)' \partial_x 
        A^{(2)} - \frac{2}{M} \partial_y B,  \label{eq:system-2-bis}
\end{eqnarray}
where use has been made of Eq.~(\ref{eq:C}).\par
%
%
To go forward in the solution of such relations, it is appropriate to make some specific
choice for $C$ and for the five integration constants $a$, $b$, $c$, $d$, $g$. Here we
consider the first subcase referred to hereabove and therefore assume $C = -q^2 < 0$.
Some comments on the two remaining subcases will be postponed until Section~4. On
taking $a = d = 1$ and $b = c = g = 0$ in (\ref{eq:C-negative}), we arrive at 
\begin{equation}
  M(x) = \sech^2 qx, \qquad A^{(1)}(x,y) = \cosh qx \sin qy, \qquad A^{(2)}(x,y) =
  - \sinh qx \cos qy.  \label{eq:M-A} 
\end{equation}
Note that with this choice, the mass is well defined on the whole real line (which would
not happen if $c$ did not vanish). Furthermore, $q$ may be treated as a deformation
parameter so that when $q \to 0$, we get back the constant-mass case $M(x) \to
1$.\par
%
%
The two conditions in (\ref{eq:system-2-bis}) can now be rewritten as
\begin{equation}
  V_{\rm eff} - V_{\rm 1,eff} = 2 q^2 \cosh^2 qx - 2 \cosh qx \csc qy\; \partial_x B
  \label{eq:system-2-ter}
\end{equation}
and
\begin{equation}
  V_{\rm eff} - V_{\rm 1,eff} = - 2 q^2 \cosh^2 qx + 2 \cosh qx \coth qx \sec qy \;
  \partial_y B,
\end{equation}
respectively. On equating their right-hand sides and setting $B(x,y) = \sin qy\;
\beta(x,y)$, we get a separable first-order partial differential equation for $\beta(x,y)$.
Its general solution leads to
\begin{equation}
  B(x,y) = (q \sinh qx + F \csch qx) \sin qy + G,  \label{eq:B}
\end{equation}
where $F$ and $G$ denote some integration constants. From (\ref{eq:system-2-ter})
and (\ref{eq:B}), we then obtain
\begin{equation}
  V_{\rm eff} - V_{\rm 1,eff} = 2qF \coth^2 qx.  \label{eq:diff-V}
\end{equation}
\par
%
%
{}Finally it remains to solve Eq.~(\ref{eq:system-3}). On taking (\ref{eq:M-A}),
(\ref{eq:B}) and (\ref{eq:diff-V}) into account, it can be easily transformed into the
partial differential equation
\begin{eqnarray}
  \lefteqn{\coth qx\; \partial_x V_{\rm eff} - \cot qy\; \partial_y V_{\rm eff}}
       \nonumber \\
  & = & - 2q [q^2 \cosh^2 qx + F(q+F) \csch^2 qx \coth^2 qx + FG \csch qx \coth^2 qx
       \csc qy].
\end{eqnarray}
We observe that such an equation is separable provided we choose a vanishing
integration constant $G$. In this case, we obtain
\begin{equation}
  V_{\rm eff}(x,y) = - q^2 \cosh^2 qx + F(q+F) \csch^2 qx + J \ln(\cosh qx \sec
        qy) + K,  \label{eq:Veff} 
\end{equation}
\begin{equation}
  V_{\rm 1,eff}(x,y) = - q^2 \cosh^2 qx + F(-q+F) \csch^2 qx + J \ln(\cosh qx \sec
        qy) + K - 2qF,  \label{eq:V1eff}
\end{equation}
in terms of two additional integration constants $J$ and $K$.\par
%
%
We conclude that Eqs.~(\ref{eq:M-A}), (\ref{eq:B}) (with $G=0$), (\ref{eq:Veff}) and
(\ref{eq:V1eff}) provide us with a particular solution to the system of partial differential
equations (\ref{eq:system-1}) -- (\ref{eq:system-3}) in the $d=2$ case. In order that this solution
be viable and useful in the context of PDM Schr\"odinger equations with a bound-state
spectrum, we have to impose some additional restrictions. This will be the purpose of
Section~3, where we shall proceed to construct and to solve  an interesting
two-dimensional PDM model.\par
%
%
\section{A NEW EXACTLY-SOLVABLE PDM MODEL IN A SEMI-INFINITE LAYER}
\setcounter{equation}{0}

Let us consider a particle moving in a semi-infinite layer of width $\pi/q$, parallel to the
$x$-axis and with impenetrable barriers at the boundaries. This means that the variables
$x$, $y$ vary in the domain
\begin{equation}
  D: \qquad 0 < x < \infty, \qquad - \frac{\pi}{2q} < y < \frac{\pi}{2q},  \label{eq:D} 
\end{equation}
and that the wavefunctions have to satisfy the conditions
\begin{equation}
  \psi(0,y) = 0, \qquad \psi\left(x, \pm \frac{\pi}{2q}\right) = 0.  \label{eq:boundary}
\end{equation}
The motion of the particle is assumed to be described by the Hamiltonian
\begin{equation}
  H^{(k)} = - \partial_x \cosh^2 qx\, \partial_x - \cosh^2 qx\, \partial_y^2 - q^2
  \cosh^2 qx + q^2 k(k-1) \csch^2 qx + q^2 v_0,  \label{eq:H}
\end{equation}
corresponding to the mass and the effective potential given in (\ref{eq:M-A}) and
(\ref{eq:Veff}), where in the latter equation we have set $J=0$ to get an acceptable
potential, as well as $F = -qk$ (with $k>0$) and $K = q^2 v_0$ (with $v_0$
arbitrary).\par
%
%
{}From the results of Section~2, it follows that $H^{(k)}$ and 
\begin{equation}
  H_1^{(k)} = - \partial_x \cosh^2 qx\, \partial_x - \cosh^2 qx\, \partial_y^2 - q^2
  \cosh^2 qx + q^2 k(k+1) \csch^2 qx + q^2 (v_0 + 2k)  \label{eq:H_1}
\end{equation}
admit the intertwining operator
\begin{equation}
  \eta^{(k)} = \sin qy(\cosh qx\, \partial_x + q \sinh qx - qk \csch qx) - \sinh qx \cos
  qy\, \partial_y  \label{eq:eta-bis}
\end{equation}
and that their respective constants of motion read
\begin{eqnarray}
  R^{(k)} & = & - \cosh^2 qx \sin^2 qy\, \partial^2_x + 2 \sinh qx \cosh qx \sin qy
       \cos qy\, \partial^2_{xy} - \sinh^2 qx \cos^2 qy\, \partial^2_y \nonumber \\
  && \mbox{} + q \sinh qx \cosh qx (1 - 4 \sin^2 qy) \partial_x
       + q \sin qy \cos qy (1 + 4 \sinh^2 qx) \partial_y \nonumber \\
  && \mbox{} + q^2 (\sinh^2 qx - \sin^2 qy - 3 \sinh^2 qx \sin^2 qy) - q^2 k (1 + 
       \csch^2 qx \sin^2 qy) \nonumber \\
  && \mbox{} + q^2 k^2 \csch^2 qx \sin^2 qy  \label{eq:R-ter} 
\end{eqnarray}
and 
\begin{equation}
  R_1^{(k)} = R^{(k)} + 2q^2 k (1 + \csch^2 qx \sin^2 qy).  \label{eq:R_1}
\end{equation}
\par
%
%
The bound-state energy spectrum and wavefunctions of $H^{(k)}$ can be determined in
two alternative ways: either by directly solving the corresponding Schr\"odinger equation,
which is separable in the variables $x$, $y$, or by taking advantage of the intertwining
operator $\eta^{(k)}$. We plan to successively review both approaches. Then, in a third
step, we will determine the relation between the two resulting sets of wavefunctions.\par
%
%
\subsection{Separation of Variables in the PDM Schr\"odinger Equation}

Let us consider the Schr\"odinger equation (\ref{eq:schrodinger-bis}) for $H^{(k)}$ with
corresponding energy eigenvalues and wavefunctions denoted by $E^{(k)}$ and
$\psi^{(k)}(x,y)$, respectively. From (\ref{eq:H}), it is clear that the operator
\begin{equation}
  L \equiv - \partial^2_y  \label{eq:L}
\end{equation}
commutes with $H^{(k)}$, so that both operators are simultaneously diagonalizable or, in
other words, Eq.~(\ref{eq:schrodinger-bis}) is separable in the variables $x$, $y$.\par
%
%
The dependence of the wavefunctions on $y$ is determined by the solution of the
eigenvalue problem for $L$ on the interval $- \pi/(2q) < y < \pi/(2q)$ with boundary
conditions coming from the second relation in (\ref{eq:boundary}). One easily finds
\begin{equation}
  L \chi_l(y) = (l+1)^2 q^2 \chi_l(q), \qquad l=0, 1, 2, \ldots,
\end{equation}
where the (normalized) eigenfunctions are given by
\begin{equation}
  \chi_l(y) = \left\{\begin{array}{ll}
      \sqrt{\frac{2q}{\pi}} \cos[(l+1)qy] & \qquad {\rm for\ } l = 0, 2, 4, \ldots \\[0.2cm]
      \sqrt{\frac{2q}{\pi}} \sin[(l+1)qy] & \qquad {\rm for\ } l = 1, 3, 5, \ldots
  \end{array} \right..  \label{eq:chi}
\end{equation}
\par
%
%
The simultaneous eigenfunctions of $H^{(k)}$ and $L$ can therefore be written as
\begin{equation}
  \psi^{(k)}_{n,l}(x,y) = \phi^{(k)}_{n,l}(x) \chi_l(y),  \label{eq:psi}
\end{equation}
where $\phi^{(k)}_{n,l}(x)$ satisfies the ordinary differential equation
\begin{eqnarray}
  \lefteqn{H^{(k)}_l \phi^{(k)}_{n,l}(x)} \nonumber \\
  & \equiv & \left(- \cosh^2 qx \frac{d^2}{dx^2} - 2q \sinh qx \cosh qx \frac{d}{dx} +
       q^2 l(l+2) \cosh^2 qx + q^2 k(k-1) \csch^2 qx \right) \nonumber \\
  && \mbox{} \times \phi^{(k)}_{n,l}(x) \nonumber \\
  & = & \left(E^{(k)}_{n,l} - q^2 v_0\right) \phi^{(k)}_{n,l}(x)  \label{eq:H-l}
\end{eqnarray}
and $n$ distinguishes between the independent solutions characterized by the same
eigenvalue of $L$.\par
%
%
By the changes of variable and of function
\begin{equation}
  e^{qx} = \tan \left[\half\left(z + \frac{\pi}{2}\right)\right], \qquad 0 < z <
  \frac{\pi}{2},
\end{equation}
\begin{equation}
  \phi^{(k)}_{n,l}[x(z)] = \sqrt{\cos z}\, \xi^{(\kappa, \lambda)}_n(z),
\end{equation}
Eq.~(\ref{eq:H-l}) can be mapped onto the (constant-mass) Schr\"odinger equation for
the two-parameter trigonometric P\"oschl-Teller potential (also termed P\"oschl-Teller I
potential)~\cite{poschl}
\begin{equation}
  \left(- \frac{d^2}{dz^2} + \kappa(\kappa-1) \csc^2 z + \lambda(\lambda-1) \sec^2 z
  \right) \xi^{(\kappa, \lambda)}_n(z) = {\cal E}^{(\kappa,\lambda)}_n \xi^{(\kappa,
  \lambda)}_n(z),
\end{equation}
where
\begin{equation}
  \kappa = k, \qquad \lambda = l + \frac{3}{2}, \qquad {\cal E}^{(\kappa,\lambda)}_n =
  \frac{1}{q^2} \left[E^{(k)}_{n,l} - q^2 v_0 + q^2 \left(k - \half\right)^2\right]. 
\end{equation}
From the known solutions of the latter~\cite{cooper, barut} in terms of Jacobi
polynomials, we obtain for the former
\begin{equation}
  E^{(k)}_{n,l} = q^2 [(2n+l+2) (2n+l+2k+1) + v_0]  \label{eq:energy}
\end{equation}
and
\begin{eqnarray}
  \phi^{(k)}_{n,l} & = & {\cal N}^{(k)}_{n,l} (\tanh qx)^k (\sech qx)^{l+2}
       P^{(k-\half, l+1)}_n(1 - 2 \tanh^2 qx), \nonumber \\
 {\cal N}^{(k)}_{n,l} & = & \left(\frac{2q (2n+l+k+\frac{3}{2}) n!\,
       \Gamma(n+l+k+\frac{3}{2})}{(n+l+1)!\, \Gamma(n+k+\half)}\right)^{1/2}. 
       \label{eq:phi}
\end{eqnarray}
\par
%
%
The functions (\ref{eq:psi}) with $\phi^{(k)}_{n,l}(x)$ and $\chi_l(y)$ given in
Eqs.~(\ref{eq:phi}) and (\ref{eq:chi}), respectively, are square integrable over $D$ with
the normalization factors as indicated. Since the PDM $M(x) = \sech^2 qx$ vanishes for
$x \to \infty$, such functions will describe bound states only if condition
(\ref{eq:wf-C2}) is satisfied in this limit. This is indeed the case because $\cosh qx
|\phi^{(k)}_{n,l}(x)|^2 \sim \exp[-(2l+3)qx]$ for $x \to \infty$.\par
%
%
We conclude that the bound-state spectrum of $H^{(k)}$ is made of an infinite number
of energy levels, given by (\ref{eq:energy}). Since $E^{(k)}_{n,l}$ only depends on the
combination
\begin{equation}
  N = 2n+l,
\end{equation}
we can rewrite it as
\begin{equation}
  E^{(k)}_N \equiv E^{(k)}_{n,l} = q^2 [(N+2)(N+2k+1) + v_0].  \label{eq:E}   
\end{equation}
The level characterized by $N$ comprises the states specified by $(n,l) = (0,N)$, $(1,
N-2)$,~\ldots, $(\frac{N}{2}, 0)$ or $(\frac{N-1}{2}, 1)$ according to whether $N$ is
even or odd. Its degeneracy is therefore given by
\begin{equation}
  \deg(N) = \left[\frac{N}{2}\right] + 1,
\end{equation}
where $[N/2]$ stands for the integer part of $N/2$.\par
%
%
Before going to the alternative approach based upon the intertwining operators
$\eta^{(k)}$ and $\eta^{(k)\dagger}$ in Subsection~3.2, it is worth making a pause for
observing that the one-dimensional Hamiltonian $H^{(k)}_l$, as defined in (\ref{eq:H-l}),
itself satisfies some intertwining relationships
\begin{equation}
  \ca^{(k)}_l H^{(k)}_l = \left(H^{(k+1)}_{l+1} + 2 q^2 k\right) \ca^{(k)}_l, \qquad
  \tca^{(k)}_l H^{(k)}_l = \left(H^{(k+1)}_{l-1} + 2 q^2 k\right) \tca^{(k)}_l
\end{equation}
with the first-order differential operators
\begin{equation}
 \ca^{(k)}_l = \cosh qx \frac{d}{dx} + q(l+2) \sinh qx - qk \csch qx, 
\end{equation}
\begin{equation}
 \tca^{(k)}_l = \cosh qx \frac{d}{dx} - ql \sinh qx - qk \csch qx.
\end{equation}
Such a property derives from the existence for $H^{(k)}_l$ of a four-way factorization
\begin{eqnarray}
  H^{(k)}_l & = & \ca^{(k)\dagger}_l \ca^{(k)}_l + c^{(k)}_l = \ca^{(k-1)}_{l-1}
       \ca^{(k-1)\dagger}_{l-1} + c^{(k)}_{l-2} \nonumber \\
  & = & \tca^{(k)\dagger}_l \tca^{(k)}_l + \tilde{c}^{(k)}_l = \tca^{(k-1)}_{l+1}
       \tca^{(k-1)\dagger}_{l+1} + \tilde{c}^{(k)}_{l+2} \nonumber \\
  c^{(k)}_l & \equiv & q^2 (l+2) (l+2k+1), \qquad \tilde{c}^{(k)}_l \equiv q^2 l
       (l-2k+1).  
\end{eqnarray}
The latter is similar to that occurring for the three-dimensional radial harmonic
oscillator~\cite{fernandez} and is directly connected with the existence of an so(4)
potential algebra for the two-parameter trigonometric P\"oschl-Teller
potential~\cite{barut}.\par
%
%
In a standard (unbroken) SUSYQM approach to $H^{(k)}_l$, one would only consider the
pair of operators $\ca^{(k)}_l$ and $\ca^{(k)\dagger}_l$. The former indeed annihilates
the ground-state wavefunction $\phi^{(k)}_{0,l}$ of $H^{(k)}_l$ and changes
$\phi^{(k)}_{n,l}$ into $\phi^{(k+1)}_{n-1,l+1}$ for $n=1$, 2,~\ldots, while the latter
generates transitions between $\phi^{(k+1)}_{n,l+1}$ and $\phi^{(k)}_{n+1,l}$ for
$n=0$, 1, 2,~\ldots. In contrast, the second pair of operators $\tca^{(k)}_l$ and
$\tca^{(k)\dagger}_l$ does not change the $n$ value as one goes this time from
$\phi^{(k)}_{n,l}$ to $\phi^{(k+1)}_{n,l-1}$ for $n=0$, 1, 2,~\ldots , or vice versa.\par
%
%
Interpreting these results in terms of the full two-dimensional Hamiltonian $H^{(k)}$ is
difficult because one is confronted with the same type of problem as that occurring in
the SUSYQM approach to the Coulomb problem~\cite{cooper}. In such a case, the
intertwining operators for the radial Hamiltonian change the angular momentum quantum
number $l$ by one unit without affecting the angular part $Y_{l,m}(\theta, \varphi)$ of
the wavefunction. As a consequence, there has been much controversy on how to apply
the SUSYQM results to atomic spectra~\cite{kostelecky}. Here too we may observe that
both intertwining operators $\ca^{(k)}_l$, $\tca^{(k)}_l$ and their Hermitian
conjugates change by one unit the quantum number $l$, now connected with the
eigenvalue of the operator $L$ defined in (\ref{eq:L}), although the corresponding
eigenfunction $\chi_l(y)$ remains the same.\par
%
%
As we plan to show in Subsection~3.3, the intertwining operators $\eta^{(k)}$ and
$\eta^{(k)\dagger}$ offer the advantage of being free from such a problem since they
act on $\chi_l(y)$ as well as on $\phi^{(k)}_{n,l}(x)$.\par
%
%
\subsection{Intertwining Operator Approach}

As established in Section~2, the operators $H^{(k)}$ and $R^{(k)}$ defined in
(\ref{eq:H}) and (\ref{eq:R-ter}), respectively, are simultaneously diagonalizable. Let us
denote their simultaneous eigenfunctions by $\Psi^{(k)}_{N,N_0}(x,y)$ and the
corresponding eigenvalues by $E^{(k)}_N$ and $r^{(k)}_{\nu}$, with $N \equiv N_0 +
\nu$. Hence
\begin{equation}
  H^{(k)} \Psi^{(k)}_{N,N_0}(x,y) = E^{(k)}_N \Psi^{(k)}_{N,N_0}(x,y), \qquad
  R^{(k)} \Psi^{(k)}_{N,N_0}(x,y) = r^{(k)}_{\nu} \Psi^{(k)}_{N,N_0}(x,y),   
  \label{eq:H-R}
\end{equation}
where $r^{(k)}_{\nu} \ge 0$ since $R^{(k)}$ is a positive-definite operator. We plan to
show herebelow that in (\ref{eq:H-R}), $N$ and $\nu$ run over all nonnegative integers,
while $N_0$ is restricted to nonnegative even integers.\par
%
%
Our first step consists in constructing the functions $\Psi^{(k)}_{N_0,N_0}(x,y)$
corresponding to $E^{(k)}_{N_0}$ and $r^{(k)}_0 = 0$. Since they belong to the
subspace spanned by the zero modes of $\eta^{(k)}$, let us inquire into such zero
modes and therefore consider the equation
\begin{equation}
  \eta^{(k)} \omega^{(k)}_s(x,y) = 0,
\end{equation}
where $s$ will serve to distinguish between independent solutions. From
(\ref{eq:eta-bis}), it follows that this first-order partial differential equation is separable.
Up to some multiplicative constant, its general solution is given by
\begin{equation}
  \omega^{(k)}_s(x,y) = (\tanh qx)^k (\sech qx)^{s+1} (\cos qy)^s,  \label{eq:zero}
\end{equation}
where $s$ is related to the separation constant. To get functions satisfying the boundary
conditions (\ref{eq:boundary}), we have to restrict $s$ to positive values. It is then easy
to check that $\omega^{(k)}_s(x,y)$ is normalizable on the domain $D$, defined in
(\ref{eq:D}), and satisfies condition (\ref{eq:wf-C2}) for $x \to \infty$.\par
%
%
On applying $H^{(k)}$ on the zero modes (\ref{eq:zero}), we obtain after a
straightforward calculation
\begin{equation}
  H^{(k)} \omega^{(k)}_s(x,y) = q^2 [(s+1)(2k+s) + v_0] \omega^{(k)}_s(x,y) - q^2
  s(s-1) \omega^{(k)}_{s-2}(x,y).  \label{eq:H-zero}
\end{equation}
Hence, amongst the zero modes $\omega^{(k)}_s(x,y)$ with $s > 0$, there is only one
that is an eigenfunction of $H^{(k)}$, namely $\omega^{(k)}_1(x,y)$:
\begin{equation}
  H^{(k)} \omega^{(k)}_1(x,y) = q^2 [2(2k+1) + v_0] \omega^{(k)}_1(x,y).
  \label{eq:H-zero-bis}
\end{equation}
As it can be checked by comparison with (\ref{eq:E}) and with (\ref{eq:chi}), (\ref{eq:psi}),
(\ref{eq:phi}), the eigenvalue and eigenfunction of $H^{(k)}$ in (\ref{eq:H-zero-bis})
coincide with the ground-state energy $E^{(k)}_0$ and ground-state wavefunction
$\psi^{(k)}_{0,0}(x,y)$, respectively. We have therefore found the (unique) solution of
(\ref{eq:H-R}) with $r^{(k)}_0 = 0$ and $N = N_0 = 0$,
\begin{equation}
  \Psi^{(k)}_{0,0}(x,y) = \psi^{(k)}_{0,0}(x,y) \propto \omega^{(k)}_1(x,y).
  \label{eq:gs}
\end{equation}
\par
%
%
We can now get the remaining solutions of (\ref{eq:H-R}) with $r^{(k)}_0 = 0$ and $N =
N_0 \ne 0$ by looking for those linear combinations of $\omega^{(k)}_s(x,y)$, $s=1$,
3,~\ldots, $N_0+1$,
\begin{equation}
  \Psi^{(k)}_{N_0,N_0}(x,y) = \sum_{s=1}^{N_0+1} \half [1 - (-1)^s] a^{(k)}_s
  \omega^{(k)}_s(x,y),  \label{eq:Psi-low}
\end{equation}
that are eigenfunctions of $H^{(k)}$. Note that in (\ref{eq:Psi-low}), $N_0$ is
necessarily restricted to even integers. On using (\ref{eq:H-zero}), we obtain a recursion
relation for the coefficients $a^{(k)}_s$ together with the eigenvalue of $H^{(k)}$,
\begin{equation}
  E^{(k)}_{N_0} = q^2 [(N_0+2)(N_0+2k+1) + v_0].
\end{equation}
The solution to the recursion relation reads
\begin{equation}
  a^{(k)}_s = (-1)^{(N_0+1-s)/2} \frac{(N_0+1)!\, \Gamma\left(\frac{N_0+s}{2}+k+1
  \right)}{2^{N_0+1-s} s! \left(\frac{N_0-s+1}{2}\right)!\,
  \Gamma\left(N_0+k+\frac{3}{2}\right)}\, a^{(k)}_{N_0+1}, \qquad s=1, 3, \ldots,
  N_0-1.
\end{equation}
The remaining multiplicative factor $a^{(k)}_{N_0+1}$ can in principle be determined
through the normalization condition of $\Psi^{(k)}_{N_0,N_0}(x,y)$ on $D$. However,
since the zero modes $\omega^{(k)}_s(x,y)$, $s=1$, 3, 5,~\ldots, form a nonorthogonal
set, this condition is not easy to work out. We shall therefore leave the construction of
fully normalized functions $\Psi^{(k)}_{N_0,N_0}(x,y)$ to Subsection~3.3, where they
will be expressed in an orthogonal basis.\par
%
%
In a second step, we can now obtain from the functions $\Psi^{(k)}_{N_0,N_0}(x,y)$ the
solutions of (\ref{eq:H-R}) associated to $\nu=1$, 2,~\ldots, and correspondingly $N =
N_0+1$, $N_0+2$,~\ldots, by a straightforward application of the intertwining relations
(\ref{eq:inter}) and (\ref{eq:inter-bis}) for the operators $\eta^{(k)}$,
$\eta^{(k)\dagger}$, $H^{(k)}$, $H^{(k)}_1$, $R^{(k)}$, and $R^{(k)}_1$. On
comparing (\ref{eq:H_1}) and (\ref{eq:R_1}) with (\ref{eq:H}) and (\ref{eq:R-ter}),
respectively, we indeed observe that
\begin{equation}
  H^{(k)}_1 = H^{(k+1)} + \epsilon^{(k)}, \qquad \epsilon^{(k)} = 2 q^2 k,
\end{equation}
\begin{equation}
  R^{(k)}_1 = R^{(k+1)} + \rho^{(k)}, \qquad \rho^{(k)} = q^2 (2k+1).
  \label{eq:R_1-bis}
\end{equation} 
In other words, $H^{(k)}$ and $R^{(k)}$ are both shape invariant~\cite{cooper}. In view
of this property, we directly obtain
\begin{equation}
  \Psi^{(k)}_{N,N_0}(x,y) = \bar{\cal N}^{(k)}_{N,N_0} \eta^{(k)\dagger}
  \eta^{(k+1)\dagger} \cdots \eta^{(k+\nu-1)\dagger} \Psi^{(k+\nu)}_{N_0,N_0}(x,y),
  \qquad \nu=1, 2, \ldots,  \label{eq:Psi}
\end{equation}
where $\bar{\cal N}^{(k)}_{N,N_0}$ is some normalization coefficient and
\begin{equation}
  E^{(k)}_N = E^{(k+\nu)}_{N_0} + \sum_{i=0}^{\nu-1} \epsilon^{(k+i)} = q^2 [(N+2)
  (N+2k+1) + v_0],  \label{eq:E-bis}
\end{equation}
\begin{equation}
  r^{(k)}_{\nu} = \sum_{i=0}^{\nu-1} \rho^{(k+i)} = q^2 \nu(\nu+2k),  \label{eq:r}
\end{equation}
for $N = N_0+\nu$. Note that the eigenvalues (\ref{eq:E-bis}) of $H^{(k)}$ coincide
with those previously found in (\ref{eq:E}) as it should be. However, the supersymmetric
approach adopted in this Subsection has the advantage of explaining the dependence of
these eigenvalues on a single quantum number $N$, which otherwise would appear to be
accidental.\par
%
%
As a final point, let us observe that the normalization coefficient $\bar{\cal
N}^{(k)}_{N,N_0}$ in (\ref{eq:Psi}) can be easily derived from the property
\begin{equation}
  \left\langle \Psi^{(k)}_{N,N_0} \left| \Psi^{(k)}_{N,N_0}\right\rangle \right. =
  \left({\frac{\bar{\cal N}^{(k)}_{N,N_0}}{\bar{\cal N}^{(k+1)}_{N-1,N_0}}}\right)^2
  \left\langle \Psi^{(k+1)}_{N-1,N_0} \left| \eta^{(k)} \eta^{(k)\dagger} \right| 
  \Psi^{(k+1)}_{N-1,N_0}\right\rangle 
\end{equation}
and Eqs.~(\ref{eq:R}), (\ref{eq:H-R}), (\ref{eq:R_1-bis}), (\ref{eq:r}). The result reads
\begin{equation}
  \bar{\cal N}^{(k)}_{N,N_0} = q^{-\nu} \left(\frac{\Gamma(2k+\nu)}{\nu!\, \Gamma
  (2k+2\nu)}\right)^{1/2}, \qquad N = N_0 + \nu.  \label{eq:n-bar}
\end{equation}
\par
%
%
\subsection{Relation between the Two Approaches}

In Subsections 3.1 and 3.2, we followed two different methods for determining the
bound-state spectrum and wavefunctions of $H^{(k)}$. The former was based on the
separability and exact solvability of the corresponding Schr\"odinger equation, while the
latter used an intertwining-operator approach and its supersymmetric interpretation.
Such methods resulted in two distinct bases for the associated Hilbert space,
$\psi^{(k)}_{n,l}(x,y)$, $n$, $l=0$, 1, 2,~\ldots, and $\Psi^{(k)}_{N, N_0}(x,y)$, $N_0
=0$, 2, 4,~\ldots, $N = N_0$, $N_0+1$, $N_0+2$,~\ldots, corresponding to definite
eigenvalues of the respective symmetry operator, $L$ or $R^{(k)}$. The aim of this
Subsection is to determine the transformation matrix between these two bases.\par
%
%
{}For such a purpose, let us first consider the action of the intertwining operators
$\eta^{(k)}$ and $\eta^{(k)\dagger}$ on the first basis wavefunctions
$\psi^{(k)}_{n,l}(x,y)$. As it is clear from the analysis of Subsection~3.2 (see
Eq.~(\ref{eq:Psi})), such operators have the property of decreasing or increasing $N =
2n+l$ by one unit, while changing $k$ into $k+1$ or vice versa. Hence, when acting on
some generic function $\psi^{(k)}_{n,l}(x,y)$, $\eta^{(k)}$, for instance, will generate
transitions to $\psi^{(k+1)}_{n-1,l+1}(x,y)$ and $\psi^{(k+1)}_{n,l-1}(x,y)$. This is
confirmed by an explicit calculation using Eqs.~(\ref{eq:eta-bis}), (\ref{eq:chi}),
(\ref{eq:psi}), (\ref{eq:phi}), and some well-known properties of Jacobi
polynomials~\cite{erdelyi}. The obtained result reads
\begin{eqnarray}
  \eta^{(k)} \psi^{(k)}_{n,l} & = & (-1)^N q \Biggl[-
       \sqrt{n\left(n+l+k+\frac{3}{2}\right)}\, \psi^{(k+1)}_{n-1,l+1} \nonumber \\
  && \mbox{} + (1 -\delta_{l,0}) \sqrt{\left(n+k+\half\right)(n+l+1)}\, 
       \psi^{(k+1)}_{n,l-1} \Biggr].  \label{eq:eta-effect}
\end{eqnarray}
Similarly, for the Hermitian conjugate operator $\eta^{(k)\dagger}$ we get
\begin{eqnarray}
  \eta^{(k)\dagger} \psi^{(k+1)}_{n,l} & = & (-1)^{N+1} q \Biggl[- (1 -\delta_{l,0}) 
       \sqrt{(n+1) \left(n+l+k+\frac{3}{2}\right)}\, \psi^{(k)}_{n+1,l-1} \nonumber \\
  && \mbox{} + \sqrt{\left(n+k+\half\right)(n+l+2)}\, \psi^{(k)}_{n,l+1} \Biggr].
       \label{eq:eta+-effect}
\end{eqnarray}
\par
%
%
We are now able to expand the members of the second basis that are zero modes of
$\eta^{(k)}$, i.e., $\Psi^{(k)}_{N_0,N_0}(x,y)$, $N_0=0$, 2, 4,~\ldots, into linear
combinations of the first basis wavefunctions $\psi^{(k)}_{n,l}(x,y)$ with $2n+l = N_0$,
\begin{equation}
  \Psi^{(k)}_{N_0,N_0}(x,y) = \sum_{n=0}^{N_0/2} X^{(k)}_{n,N_0-2n}
  \psi^{(k)}_{n,N_0-2n}(x,y), \qquad N_0=0, 2, 4, \ldots,  \label{eq:Psi-0}  
\end{equation}
where $X^{(k)}_{n,N_0-2n}$ are some coefficients to be determined. On applying
$\eta^{(k)}$ on both sides of this equation and using the facts that the left-hand side
vanishes while Eq.~(\ref{eq:eta-effect}) allows us to calculate the right-hand one, we
indeed obtain a recursion relation for the coefficients $X^{(k)}_{n,N_0-2n}$, whose
solution is given by
\begin{equation}
  X^{(k)}_{n,N_0-2n} = \left(\frac{\left(\frac{N_0}{2}\right)!
  \left(\frac{N_0}{2}+1\right)!\, \Gamma\left(N_0-n+k+\frac{3}{2}\right)
  \Gamma\left(n+k+\half\right)}{n!\, (N_0-n+1)!\,
  \Gamma\left(\frac{N_0+3}{2} +k\right) \Gamma\left(\frac{N_0+1}{2} +k\right)} 
  \right)^{1/2} X^{(k)}_{N_0/2,0}.  \label{eq:X-X_0}
\end{equation}
As shown in the Appendix (see Eq.~(\ref{eq:X-0})), the remaining multiplicative factor
$X^{(k)}_{N_0/2,0}$ can be computed from the normalization condition of $\Psi^{(k)}_{N_0,
N_0}(x,y)$. The final result for $X^{(k)}_{n, N_0-2n}$ can then be written as
\begin{equation}
  X^{(k)}_{n,N_0-2n} = \left(\frac{(N_0+1)!\, \Gamma(k+1)
  \Gamma\left(n+k+\half\right) \Gamma\left(N_0-n+k+\frac{3}{2}\right)} {2^{N_0} n!\,
  (N_0-n+1)!\, \Gamma\left(\frac{N_0}{2}+k+1\right) \Gamma\left(k+\half\right)
  \Gamma\left(\frac{N_0+3}{2}+k\right)}\right)^{1/2}.  \label{eq:X}
\end{equation}
For the first few $N_0$ values, Eq.~(\ref{eq:Psi-0}) reads
\begin{equation}
  \Psi^{(k)}_{0,0} = \psi^{(k)}_{0,0},  \label{eq:gs-bis}
\end{equation}
\begin{equation}
  \Psi^{(k)}_{2,2} = \frac{1}{2\sqrt{k+1}} \left(\sqrt{k+\frac{5}{2}}\, \psi^{(k)}_{0,2}
  + \sqrt{3\left(k+\half\right)}\, \psi^{(k)}_{1,0}\right),  \label{eq:Psi-2}
\end{equation}
\begin{eqnarray}
  \Psi^{(k)}_{4,4} & = & \frac{1}{4\sqrt{(k+1)(k+2)}}
      \Biggl(\sqrt{\left(k+\frac{7}{2}\right) \left(k+\frac{9}{2}\right)}\, \psi^{(k)}_{0,4}
      + \sqrt{5\left(k+\half\right) \left(k+ \frac{7}{2}\right)}\, \psi^{(k)}_{1,2}
      \nonumber \\
  && \mbox{} + \sqrt{10\left(k+\half\right) \left(k+ \frac{3}{2}\right)}\,
      \psi^{(k)}_{2,0} \Biggr).  \label{eq:Psi-4} 
\end{eqnarray}
Note that Eq.~(\ref{eq:gs-bis}) coincides with (\ref{eq:gs}) as it shoud be.\par
%
%
To be able to extend expansion (\ref{eq:Psi-0}) to all members
$\Psi^{(k)}_{N,N_0}(x,y)$ of the second basis, defined in (\ref{eq:Psi}), we have first to
iterate Eq.~(\ref{eq:eta+-effect}) to determine the action of $\nu$ operators of type
$\eta^{(k)\dagger}$ on the first basis wavefunctions:
\begin{equation}
  \eta^{(k)\dagger} \eta^{(k+1)\dagger} \cdots \eta^{(k+\nu-1)\dagger}
  \psi^{(k+\nu)}_{n,l} = \sum_{\mu=0}^{\nu} Y^{(k)}_{n, l; n+\mu, l+\nu-2\mu}
  \psi^{(k)}_{n+\mu, l+\nu-2\mu}.\label{eq:prod-eta}
\end{equation}
Here the coefficients are given by
\begin{eqnarray}
  && Y^{(k)}_{n, l; n+\mu, l+\nu-2\mu} = (-1)^{\nu \left(N +
       \frac{\nu+1}{2}\right) + \mu} q^{\nu} x_{\mu}(l,\nu) \nonumber \\
  && \times \left(\frac{(n+\mu)!\, (n+l+\nu-\mu+1)!\, \Gamma\left(n+\nu+k
       +\half\right) \Gamma\left(n+l+\nu+k+\frac{3}{2}\right)}{n!\, (n+l+1)!\,
       \Gamma\left(n+\mu+k+\half\right)
       \Gamma\left(n+l+\nu-\mu+k+\frac{3}{2}\right)}\right)^{1/2}  \label{eq:Y}  
\end{eqnarray}
where $x_{\mu}(l,\nu)$ is a combinatorial factor giving the number of ways one can
go from $l$ to $l+\nu-2\mu$ in $\nu$ steps so that in each step $l$ increases or
decreases by one unit without reaching negative values. This means in particular that
\begin{equation}
  x_{\mu}(l,\nu) = 0 \qquad {\rm if\ } \mu > \left[\frac{l+\nu}{2}\right].  \label{eq:combi-0}
\end{equation}
Hence the summation on the right-hand side of (\ref{eq:prod-eta}) actually runs from
0 to $\min\left(\nu,\left[\frac{l+\nu}{2}\right]\right)$.\par
%
%
{}For $\nu=1$, Eq.~(\ref{eq:prod-eta}) reduces to Eq.~(\ref{eq:eta+-effect}) since in
such a case
\begin{equation}
  x_0(l,1) = 1, \qquad x_1(l,1) = 1 - \delta_{l,0}.  \label{eq:combi-1}
\end{equation}
For higher $\nu$ values, it can be easily proved by induction over $\nu$ on taking the
relations
\begin{eqnarray}
  x_0(l,\nu) & = & x_0(l,\nu-1), \nonumber \\
  x_{\mu}(l,\nu) & = & \left(1 - \delta_{\mu, (l+\nu+1)/2}\right) x_{\mu-1}(l,\nu-1)
       + x_{\mu}(l,\nu-1), \qquad \mu=1, 2, \ldots, \nu-1, \nonumber \\
  x_{\nu}(l,\nu) & = & \left(1 - \delta_{l,\nu-1}\right) x_{\nu-1}(l,\nu-1)  \label{eq:combi-syst}
\end{eqnarray}
into account. Some examples of explicit solutions to this set of recursion relations
are provided in the Appendix.\par
%
%
{}Finally, it remains to combine Eqs.~(\ref{eq:Psi}), (\ref{eq:Psi-0}), and
(\ref{eq:prod-eta}) to obtain the searched for expansion
\begin{equation}
  \Psi^{(k)}_{N,N_0} = \sum_{n=0}^{[N/2]} Z^{(k)}_{N_0;n,N-2n} \psi^{(k)}_{n,N-2n},
\end{equation}
where
\begin{equation}
  Z^{(k)}_{N_0;n,N-2n} = \bar{\cal{N}}^{(k)}_{N,N_0} \sum_{n'=n'_{\rm min}}^{n'_{\rm
  max}}  X^{(k+\nu)}_{n',N_0-2n'} Y^{(k)}_{n',N_0-2n';n,N-2n}.  \label{eq:Z}  
\end{equation}
Here $n'_{\rm min} \equiv \max(0,n+N_0-N)$, $n'_{\rm max} \equiv
\min\left(\frac{N_0}{2},n\right)$ and the three factors on the right-hand side are
given by Eqs.~(\ref{eq:n-bar}), (\ref{eq:X}) and (\ref{eq:Y}), respectively.\par
%
%
{}For those wavefunctions of the second basis $\Psi^{(k)}_{N,0}(x,y)$ that can be
reached from the ground state (\ref{eq:gs-bis}) by applying $N$ operators of type
$\eta^{(k)\dagger}$, Eq.~(\ref{eq:Z}) can be written explicitly as
\begin{eqnarray}
  && Z^{(k)}_{0;n,N-2n} = (-1)^{\half N(N+3)+n} (N-2n+1) \nonumber \\
  && \times \left(\frac{N!\, \Gamma\left(\frac{N}{2}+k\right)
       \Gamma\left(\frac{N+1}{2}+k\right) \Gamma\left(N+k+\frac{3}{2}\right)}
       {2^N n!\, (N-n+1)!\, \Gamma(N+k) \Gamma\left(n+k+\half\right)
       \Gamma\left(N-n+k+\frac{3}{2}\right)}\right)^{1/2}
\end{eqnarray}
on using Eq.~(\ref{eq:combi}) for the combinatorial factor $x_{\mu}(0,\nu)$. So,
for instance,
\begin{equation}
  \Psi^{(k)}_{1,0} = \psi^{(k)}_{0,1},
\end{equation}
\begin{equation}
  \Psi^{(k)}_{2,0} = \frac{1}{2\sqrt{k+1}} \left(- \sqrt{3\left(k+\half\right)}\,   
  \psi^{(k)}_{0,2} + \sqrt{k+\frac{5}{2}}\, \psi^{(k)}_{1,0}\right),  \label{eq:Psi-2-bis} 
\end{equation}
\begin{equation}
  \Psi^{(k)}_{3,0} = \frac{1}{\sqrt{2(k+2)}} \left(- \sqrt{k+\half}\, \psi^{(k)}_{0,3} +
  \sqrt{k+\frac{7}{2}}\, \psi^{(k)}_{1,1}\right), 
\end{equation}
\begin{eqnarray}
  \Psi^{(k)}_{4,0} & = & \frac{1}{4\sqrt{(k+2)(k+3)}} \Biggl(
       \sqrt{5\left(k+\half\right) \left(k+\frac{3}{2}\right)}\, \psi^{(k)}_{0,4} - 3
       \sqrt{\left(k+\frac{3}{2}\right) \left(k+\frac{9}{2}\right)}\, \psi^{(k)}_{1,2}
       \nonumber \\
  && \mbox{} + \sqrt{2\left(k+\frac{7}{2}\right) \left(k+\frac{9}{2}\right)}\,
       \psi^{(k)}_{2,0}\Biggr).  \label{eq:Psi-4-bis}  
\end{eqnarray}      
As it can be checked, the functions (\ref{eq:Psi-2}) and (\ref{eq:Psi-2-bis}), as well
as (\ref{eq:Psi-4}) and (\ref{eq:Psi-4-bis}), are orthogonal.\par
%
%
\section{FINAL REMARKS}

In this paper, we have developed a general framework for analyzing $d$-dimensional PDM
Hamiltonian pairs $(H, H_1)$ admitting a first-order intertwining operator $\eta$. We
have established that for $d \ge 2$, there always exists another pair of intertwined
second-order partial differential operators $(R, R_1)$ associated with the same operator
$\eta$ and such that $R$ (resp.\ $R_1$) commutes with $H$ (resp.\ $H_1$). We have
shown that in a SUSYQM context based on an sl(1/1) superalgebra, $R$ and $R_1$ can
be interpreted as SUSY partners, while $H$ and $H_1$ are related to the Casimir operator
of a larger gl(1/1) superalgebra. Furthermore, we have derived a system of partial
differential equations to be satisfied by the mass, the functions appearing in the definition
of $\eta$ and the effective potentials contained in $H$ and $H_1$.\par
%
%
Considering this system in more detail in the two-dimensional case for a PDM depending on
a single variable, we have proved that the mass may be of a hyperbolic, trigonometric or
rational nature. We have then chosen the class of hyperbolic PDM's and, more specifically, a
$\sech^2$-mass, which has proved very useful in the context of exactly-solvable
one-dimensional problems~\cite{bagchi04, bagchi05b}. Under this assumption, we have
obtained the general solution to the system of partial differential equations when some
integration constant ($G$ in (\ref{eq:B})) vanishes.\par
%
%
{}From such a solution, we have built a physically-relevant model depicting the motion of a
particle in a semi-infinite layer. The corresponding Hamiltonian $H$ has two noncommuting
integrals of motion, $L$ and $R$. Diagonalizing $H$ and $L$ simultaneously amounts to
separating the Schr\"odinger equation into two exactly-solvable differential equations. On
proceeding along these lines, we have obtained a bound-state spectrum made of an
infinite number of levels exhibiting some finite degeneracies. Choosing next to
simultanously diagonalize the shape-invariant operators $H$ and $R$, we have taken
advantage of poweful SUSYQM techniques to obtain their respective spectrum and
eigenfunctions in a straightforward way. The latter approach has had two interesting
outcomes. First, it has shed some light on the origin of the `accidental' degeneracies
observed in the Hamiltonian spectrum derived in the former approach. Second, it has provided
us with a partnership between the full two-dimensional Hamiltonians $H$ and $H_1$, free
from the interpretation problems that plague the more traditional description based on
one-dimensional intertwining operators~\cite{kostelecky}.\par
%
%
Had we chosen the class (\ref{eq:C-positive}) of trigonometric PDM's instead of that of
hyperbolic ones, a solution to the system of partial differential equations similar to that
obtained in Section~2 would have emerged. The counterpart of the model considered in
Section~3, now depicting a particle moving in a rectangular box, however turns out to be
devoid of bound states. Constructing some physically-relevant model in such a case
therefore remains under study.\par
%
%
{}Finally, in view of an equivalence between Schr\"odinger equations involving a PDM and
those in a curved space~\cite{cq}, selecting the class (\ref{eq:C-null}) of rational PDM's
and, more specifically, a $1/x^2$-mass would have led to some results directly connected
with those recently obtained for intertwined Hamiltonians in Poincar\'e half plane
($AdS_2$)~\cite{samani}.\par
%
%
{}From a physical viewpoint, our model might find applications in the study of quantum wires with
an abrupt termination in an environment that can be modelled by a dependence of the carrier
effective mass on the position. The presence of bound states may be compared to that observed in
a quantum channel whenever the uniformity is broken, for instance by a quantum dot or a bend
(see, e.g., \cite{olendski} and references quoted therein).\par
%
%
Some interesting open problems for future work are the extensions of the present study
to higher-dimensional models and to second-order intertwining operators (for a recent
review on the latter see, e.g.,~\cite{ioffe}).\par
%
%
\section*{APPENDIX}

\renewcommand{\theequation}{A.\arabic{equation}}
\setcounter{equation}{0}

In this Appendix, we prove some results used in the calculation of the transformation
matrix between the two bases $\{\psi^{(k)}_{n,l}\}$ and $\{\Psi^{(k)}_{N,N_0}\}$ in
Subsection~3.3.\par
%
%
Let us first consider the computation of the coefficient $X^{(k)}_{N_0/2,0}$ in 
expansion (\ref{eq:Psi-0}). On using Eq.~(\ref{eq:X-X_0}), the normalization condition
$\sum_{n=0}^{N_0/2} \left(X^{(k)}_{n,N_0-2n}\right)^2 = 1$ of
$\Psi^{(k)}_{N_0,N_0}(x,y)$ can be written in terms of
\begin{equation}
  S^{(k)}_{N_0} = \sum_{n=0}^{N_0/2} \frac{\Gamma\left(N_0-n+k+\frac{3}{2}\right)
  \Gamma\left(n+k+\half\right)}{n!\, (N_0-n+1)!} = \frac{1}{2} \sum_{n=0}^{N_0+1} 
  \frac{\Gamma\left(N_0-n+k+\frac{3}{2}\right)
  \Gamma\left(n+k+\half\right)}{n!\, (N_0-n+1)!}.
\end{equation}
The identity $(1+x)^{-\left(k+\half\right)} (1+x)^{-\left(k+\half\right)} =
(1+x)^{-(2k+1)}$, combined with the binomial theorem, leads to the closed form
\begin{equation}
  S^{(k)}_{N_0} = \frac{\Gamma(N_0+2k+2) \Gamma^2\left(k+\half\right)}{2 (N_0+1)!\,
  \Gamma(2k+1)}.
\end{equation}
From this result and some elementary properties of the gamma function~\cite{erdelyi}, we
obtain
\begin{equation}
  X^{(k)}_{N_0/2,0} = \left(\frac{(N_0+1)!\, \Gamma\left(\frac{N_0+1}{2}+k\right)
  \Gamma(k+1)}{2^{N_0} \left(\frac{N_0}{2}\right)!\, \left(\frac{N_0}{2}+1\right)!\,
  \Gamma\left(k+\half\right) \Gamma\left(\frac{N_0}{2}+k+1\right)}\right)^{1/2}. 
  \label{eq:X-0}
\end{equation}
\par
%
%
Let us next provide some solutions to the set (\ref{eq:combi-syst}) of recursion relations
for the combinatorial factor $x_{\mu}(l,\nu)$ with the starting values given in
(\ref{eq:combi-1}). The simplest case corresponds to $l\ge \nu$, because in none of the
$\nu$ steps from $l$ to $l+\nu-2\mu$ can a negative $l$ value be reached. From
(\ref{eq:combi-0}), it follows that $x_{\mu}(l,\nu)$ is nonvanishing for $\mu=0$,
1,~\ldots, $\nu$. Furthermore, Eq.~(\ref{eq:combi-syst}) reduces to   
\begin{eqnarray}
  x_0(l,\nu) & = & x_0(l,\nu-1), \nonumber \\
  x_{\mu}(l,\nu) & = & x_{\mu-1}(l,\nu-1) + x_{\mu}(l,\nu-1), \qquad \mu=1, 2, \ldots,
         \nu-1, \nonumber \\
  x_{\nu}(l,\nu) & = & x_{\nu-1}(l,\nu-1),
\end{eqnarray}
so that $x_{\mu}(l,\nu)$ is simply given by a binomial coefficient
\begin{equation}
  x_{\mu}(l,\nu) = \left( \begin{array}{c}
      \nu \\
      \mu
  \end{array}\right), \qquad \mu=0, 1, \ldots, \nu, \qquad {\rm if\ } l \ge \nu.
\end{equation}
\par
%
%
In contrast, the value $l=0$ maximizes the number of possible encounters with negative
$l$ values during the $\nu$ steps. The nonvanishing values of $x_{\mu}(l,\nu)$ now
correspond to $\mu=0$, 1,~\ldots, $[\frac{\nu}{2}]$ and satisfy the relations
\begin{eqnarray}
  x_0(0,\nu) & = & x_0(0,\nu-1), \nonumber \\
  x_{\mu}(0,\nu) & = & x_{\mu-1}(0,\nu-1) + x_{\mu}(0,\nu-1), \qquad \mu=1, 2, \ldots,
         \frac{\nu}{2}-1, \nonumber \\
  x_{\frac{\nu}{2}}(0,\nu) & = & x_{\frac{\nu}{2}-1}(0,\nu-1),
\end{eqnarray}
or
\begin{eqnarray}
  x_0(0,\nu) & = & x_0(0,\nu-1), \nonumber \\
  x_{\mu}(0,\nu) & = & x_{\mu-1}(0,\nu-1) + x_{\mu}(0,\nu-1), \qquad \mu=1, 2, \ldots,
         \frac{\nu-1}{2}, 
\end{eqnarray}
for even and odd values of $\nu$, respectively. In both cases, the solution reads
\begin{equation}
  x_{\mu}(0,\nu) = \left\{\begin{array}{ll}
     \frac{\nu!\, (\nu-2\mu+1)}{\mu!\, (\nu-\mu+1)!} & \qquad\mu=0, 1, \ldots,
          \left[\frac{\nu}{2}\right]  \\[0.2cm]
     0 & \qquad \mu = \left[\frac{\nu}{2}\right]+1, \left[\frac{\nu}{2}\right] +2, \ldots,
          \nu 
     \end{array}\right..  \label{eq:combi}
\end{equation}
\par
%
%
\section*{ACKNOWLEDGMENT}

The author would like to thank V.\ M.\ Tkachuk for an interesting discussion. She is a Research
Director of the National Fund for Scientific Research (FNRS), Belgium.\par
%
%

\end{document}